\def \yskip{\penalty-50\vskip3pt plus 3pt minus 2pt}
\def \reference{\par \yskip \noindent \hangindent .4in \hangafter 1}
\def \abc#1#2#3#4 {\reference#1, {\sl#2}, {\bf#3}, #4}
\def \blank {\lower 5pt\hbox to 0.75in{\hrulefill}}

\def \cm{\rm{cm}}

\def \s{\rm{s}}
\def \km{\rm{km}}
\def \lae{\mathrel{<\kern-1.0em\lower0.9ex\hbox{$\sim$}}}
\def \gae{\mathrel{>\kern-1.0em\lower0.9ex\hbox{$\sim$}}}
\def \yskip{\penalty-50\vskip3pt plus 3pt minus 2pt}
\def \reference{\par \yskip \noindent \hangindent .4in \hangafter 1}
\def \abc#1#2#3#4 {\reference#1, {\sl#2}, {\bf#3}, #4}
\def \blank {\lower 5pt\hbox to 0.75in{\hrulefill}}

\def \cm{\rm{cm}}

\def \s{~\rm{s}}
\def \km{~\rm{km}}

\def \erg{\rm{erg}}
\def \mjy{\rm{mJy}}
\def \lae{\mathrel{<\kern-1.0em\lower0.9ex\hbox{$\sim$}}}
\def \gae{\mathrel{>\kern-1.0em\lower0.9ex\hbox{$\sim$}}}

\documentstyle[11pt,aasms4]{article}

\begin{document}
\small

\setcounter{page}{1}

\begin{center}
\bf NONTHERMAL RADIO EMISSION FROM PLANETARY NEBULAE
\end{center}

\vspace*{0.5cm}

\begin{center}
Ruth Dgani \\
knill@astro.as.utexas.edu\\
Department of Astronomy, The University of Texas\\
Austin, TX 78712\\
and\\
Noam Soker\\
soker@physics.technion.ac.il \\
Department of Physics, University of Haifa at Oranim\\
Tivon 36006, Israel

\end{center}

\clearpage 

\begin{center}
\bf ABSTRACT
\end{center}

 In a recent analysis of the radio emission from the planetary nebula
A30, Dgani, Evans \& White (1998) claim that the emission,  located 
in the inner region, is  probably dominated by nonthermal emission. 
 We propose a model to explain this.
 We assume that the fast wind, blown by the central star of A30 carries 
a very weak magnetic field.
The interaction of this wind with a cluster of dense condensations traps
 the magnetic field lines for a long time and stretches them, 
leading to a strong magnetic field. 
 If relativistic particles are formed as the fast
wind is shocked, then the enhanced magnetic field will 
result in nonthermal radio emission.
 The typical nonthermal radio flux at 1 GHz can be up to several 
milli-Jansky.
 In order to detect the nonthermal emission, the emitting region 
should be spatially resolved from the main optical nebula.
 We list other planetary nebulae which may possess nonthermal 
radio emission.

\bigskip

{\it Subject heading:}
 ISM: magnetic fields
$-$ ISM: kinematics and dynamics
$-$ Radio continuum: ISM
$-$ Planetary nebulae: general
$-$ Planetary nebulae: individual (A30)
\clearpage 

\section{INTRODUCTION}

The possibility of magnetic fields carried by the fast winds in 
planetary nebulae (PNs; Chevalier and Luo 1994; hereafter CL; 
Chevalier 1995;  Garcia-Segura 1997) raises the question of whether the 
interaction between this wind and the previously ejected slow wind gives rise 
to particle acceleration, and hence nonthermal emission from relativistic 
electrons (Chevalier 1995). 
 Chevalier notes that the major problem with detecting the 
nonthermal emission is the strong thermal free-free emission emanating 
from the dense shell.
 In many PNs, however, knots and tails are present relatively close to 
the central star, where both the ram pressure of the fast wind and the 
magnetic field are higher, increasing the nonthermal emission.
 In the hydrogen deficient PNs A30 and A78, for example, knots with tails
appear in a disk structure close to the central star 
(Borkowski, Harrington, \& Tsvetanov 1995).
These nebulae are believed to have had a helium flash late in their 
evolution, and are known to possess fast winds, $v_w \sim 4000 $ km s$^{-1}$
(Leuenhagen, Koesterke \& Hamann 1993 ). 
The knots are believed to be photo evaporating globules (Dyson 1968; 
Kahn 1969; Borkowski {\it et al.} 1995).
 The HST images of the knots in the central region of A30 
(Borkowski {\it et al.} 1995) show that they are being disintegrated 
and accelerated by the dilute shocked fast wind from the central 
star, giving rise to Rayleigh-Taylor instability.
In addition, hydrogen deficient nebulae have lower thermal emission (see
discussion in 
Dgani, Evans \& White 1998), allowing the nonthermal component to be detected.

  Jones, Kang \& Tregillis (1994) have simulated the evolution of dense
gas clouds moving supersonically through a magnetized low density medium. 
They took a passive magnetic field and found that the synchrotron emission 
from relativistic electrons increase rapidly when the cloud begins to 
fragment under  Rayleigh-Taylor instability. 
 The synchrotron emission is dominated by field enhancement through  
stretching in the shear turbulent layers formed  between
and behind the fragments.
Jones, Ryu \& Tregillis (1996) performed full two dimensional MHD
calculations and found that when the ambient field is transverse to 
the flow fast amplification of the field occurs also at the nose of the 
cloud, and not only in the turbulent regions.
   
Dgani {\it et al.} (1998)  present a  radio map of the central
region of A30. The map shows an extended source  of about 0.15
mJy stretching from the central star
to  a point 3'' to the west.
Based on  a comparison to the optical line emission from that 
region, they claim that the source is  probably nonthermal.
 Motivated by this recent finding, we apply  the  theoretical results 
of Jones {\it et al.} (1994; 1996) to the interaction of a magnetic fast wind 
with knots in PNs. 
 The amplification of the magnetic field is calculated in $\S 2$,
while in $\S 3$ we estimate the nonthermal radio emission from the 
interaction region.
 In $\S 4$ we summarize, and list PNs, other than A30 
(Dgani {\it et al.} 1998), which we expect to have nonthermal radio
emission.

\section {INTERACTION OF DENSE KNOTS WITH A MAGNETIZED WIND}

 Like CL, we examine a slowly rotating central star, for which the 
tangential component of the magnetic field dominates far from the star.
For a field line that starts radially from the stellar equator, the downstream 
tangential component of the field at $r \gg r_s$, where $r$ is the distance 
from the central star and $r_s$ its radius,  is given by 
\begin{eqnarray}
B=B_s {v_{\rm {rot}}\over v_w}{r_s\over r},
\end{eqnarray}
where $v_{\rm {rot}}$ is the equatorial velocity of the rotating star,
$v_w$ is the fast wind terminal velocity, and $B_s$ is the 
(radial) magnetic field on the stellar surface. 
Since both the magnetic pressure $B^2/8\pi$ and the ram pressure 
$\rho v_w^2$ drop as $r^{-2}$ before hitting the shock wave, it is useful
to define the parameter (CL): 
\begin{eqnarray}
\sigma={B^2\over 4\pi \rho v_w^2}={B_s^2 r_s^2\over \dot M v_w}
\left({v_{rot}\over v_w}\right)^2 ,
\end{eqnarray}
where $\dot M$ is the mass loss rate to the fast wind and 
$\rho=\dot M (4\pi r^2 v_w)^{-1}$ is the density in the wind.
$\sigma$ is the ratio of field energy density to kinetic energy density
 in the wind.
 CL find that the magnetic field shapes the nebula for $\sigma \gae 10^{-4}$.
 From equation (2) we can write the magnetic field in the
equatorial plane in the  free wind as
\begin{eqnarray}
B=3.5 \times 10^{-7}
\left({\sigma\over 10^{-8}}\right)^{1\over 2}
\left({{\dot M}\over{ 5 \times 10^{-8} M_{\odot}~yr^{-1}}}\right)^{1\over 2}
\left({v_w\over 4000 km ~s^{-1}}\right)^{1\over 2}
\left({r\over 10^{16}\cm}\right)^{-1} G  .
\end{eqnarray} 
The mass loss rate is scaled according to the value 
Borkowski {\it et al.} (1995) used for A30.

Consider first a single globule with radius $R_g$.
A supersonic flow incident on it will form a bow shock at its nose, 
a turbulent shear layer around its sides and a turbulent wake behind it.
The magnetic field is expected to be enhanced in these regions
because of line stretching (e.g., Jones {\it et al.} 1994, 1996). 
 As  shown in the extensive numerical simulations of
Jones {\it et al.} (1996),  when the magnetic field lines are transverse to 
the flow the magnetic pressure on the nose of the clump reaches equipartition 
with the ram pressure, even if the initial magnetic field is quite weak.
 Using their analysis for the numerical results, we
can estimate the growth rate of the magnetic field by the following
consideration (their equations 8-10)

  Jones {\it et al.} (1996, eq. 8) start with the ideal MHD magnetic 
induction equation
\begin{eqnarray}
{d ln(B/\rho)\over dt}={B\cdot[(B\cdot\nabla)u]\over B^2} .
\end{eqnarray}
$u$, $\rho$ and $B$ are the velocity, the density and the magnetic field
in the flow respectively.
 Taking $(B\cdot\nabla)u\sim B u/R_g$, equation (4) becomes
\begin{eqnarray}
{d ln(B/\rho)\over dt}={u\over R_g} .
\end{eqnarray}
 The last equation gives exponential growth to the magnetic field as 
long as the magnetic field lines stay near the stagnation point.
Jones {\it et al.} (1996) estimate that at the stagnation point 
the magnetic pressure will rise exponentially until the bullet
is crushed or until the magnetic pressure  equals the ram pressure. 
However, since they performed 2D calculations, they neglected the 
sliding of the magnetic field line around the bullet. 
Although they claim  irregularities in the 3D spherical bullet 
will capture the field lines and prevent sliding, we believe
that for a flow with very large $\beta=P_{\rm {th}}/P_B$, the
field lines will be carried around the globule in a time $\lae R_g/u$,
much shorter than the bullet's crushing time.
 Here $P_{\rm {th}}$ and $P_B$ are thermal and magnetic pressures, 
respectively.

 Other sites of magnetic field enhancement are the turbulent mixing  
layer at the sides of the globule, and its turbulent wake.
 These are more promising sites, mainly because the clumps (globules) appear in groups, 
and turbulent flows dominate the region. 

In here, we 
consider the flow around and inside an ensemble of clumps,
occupying a region of size $R_{\rm {ens}}$.
 The typical distance between the clumps is of the order of their size 
$R_g$.
 This case has not been studied numerically, but we can apply an analytical 
approach to estimate the magnetic field enhancement.
 A turbulent flow will form between the globules, where the magnetic field 
can be amplified.  
 There, the change of the wind velocity $u$ along a field line
depends on the size of the  mixing layer  $ l\sim R_g$.
That is  $(B\cdot\nabla)u\sim B u/R_g$.
The time a field line spends in the turbulent region is the time it takes
the flow to pass the group:   $R_{ens}/u$.
Substituting these values in the induction equation (4),
and integrating over the time the field lines spend inside the group 
of globules, gives the relative amplification
\begin{eqnarray}
 B/\rho \sim  \exp {R_{ens}\over R_g} .
\end{eqnarray}
 For $R_g\sim0.1 R_{ens}$, for example, there is a large  
amplification factor $\sim 10^4$. 
 Therefore, the magnetic field will reach equipartition even if its
initial value is $\sim 10^{-4}$ that of equipartition, i.e.,
$\sigma \sim 10^{-8}$ (eq. 2). 
  In the model of CL and Garcia-Segura (1997) the post shock magnetic 
field grows as $B\sim r$, which gives a factor of $\sim 10-100$ for the 
magnetic field and $\sim 100-10^{4}$ for the magnetic pressure. 
  Since CL assume that the magnetic field shapes the PN globally, they have 
to assume that the initial magnetic pressure is $ \gae 10^{-4} \rho v^2$, 
where $\rho v^2$ is the fast wind's ram  pressure. 
 This means $\sigma > 10^{-4}$.
 In the flow pattern assumed here, the localized turbulence results 
in a much larger increase in the magnetic field. 
 Therefore, to achieve equipartition between the magnetic pressure 
and the post shock thermal pressure, a value of $\sigma \gae 10^{-8}$ 
is sufficient. 

\section {NONTHERMAL RADIO EMISSION}
 
 The upper limit for the enhanced magnetic field is the equipartition value, 
found from  the relation $P_B=\rho v_w^2$,  
\begin{eqnarray}
B_{max} \simeq 5\times 10^{-3}
 \left({{\dot M}\over{ 5\times10^{-8} M_{\odot}~yr^{-1}}}\right)^{1\over 2}
  \left({v_w\over 4000 \km ~\s^{-1}}\right)^{1\over 2}
  \left({d\over 10^{16}\cm}\right)^{-1} G, 
\end{eqnarray}
where $d$ is the distance between the central star and the group of globules.
 If the relativistic electrons are produced in the bow shock and then
carried downstream to the magnetized region, they can emit
synchrotron radiation.
We assume equipartition between the magnetic field and relativistic 
particles (electrons and nuclei).
The radio luminosity between $\nu_1 $ and $\nu_2$   can then
be estimated if the spectral index $\alpha$, (i.e., the flux
$F_\nu \propto \nu^{-\alpha}$), of the radiation is known
(cf. Pacholczyk 1970 Page 171, Eq. 7.14).
The radio luminosity is given by
\begin{eqnarray}
L={B^{7/2}\; \Phi\; R_{ens}^3 \over 4.5\; (1+k)\; 
c_{12}(\alpha,\nu_1,\nu_2)} ,
\end{eqnarray}
where $\Phi$ is the fraction of the ensemble volume occupied by
magnetic field and relativistic particles, k is the ratio of the 
heavy particle energy to the electron energy, and
$c_{12}(\alpha,\nu_1,\nu_2)$ is tabulated in appendix 2 of Pacholczyk.

Taking   $\nu_1 =10^7$ Hz and $\nu_2=10^{11}$ Hz,  $\alpha=0.75$,  
for which $c_{12}=5\times 10^7$ and $k=100$, (Chevalier 1977;1995) 
we obtain
\begin{eqnarray}
L \simeq 4.9\times 10^{28}\Phi\left({R_{ens}\over 5\times 10^{15}\cm}\right)^3
\left({B\over 5\times 10^{-3} G}\right)^{7/2}
\erg \s^{-1} .
\end{eqnarray}
 The radio flux at 1GHz can  now be estimated:
\begin{eqnarray}
F_{\nu9} \simeq 3.5 \Phi\left({R_{ens}\over 5\times 10^{15}\cm}\right)^3
\left({B\over 5\times 10^{-3} G}\right)^{7/2}
\left({D\over  kpc}\right)^{-2} \mjy
\end{eqnarray}
where D is the distance to the source, and
$ 1 \mjy = 10^{-29}$ W m$^{-2}$ Hz$^{-1}$.
A few mJy is much less than the observed fluxes from the shells of most
catalogued PNs (Zijlstra, Pottasch, \& Bignell 1989).
 However, the emission we  predict is far from the PN shell, 
near the central star, and  will not be overwhelmed 
by thermal emission.
Chevalier (1995) discussed the case in which  nonthermal electrons are 
produced in the fast wind termination shock,
and then carried  downstream to the magnetized region near the shell.
In this case the nonthermal emission region neighbors
the thermal emission region and cannot be separated observationally. 
Therefore, Chevalier  concluded that a search for radio synchrotron emission 
from planetary nebulae would involve  a frequency lower than 1GHz.

Dgani {\it et al.} (1998) show  an 8 GHz radio map of 
the hydrogen deficient PN A30.
 They present a source of $\sim 0.1$ mJy which  extends
to about 3$^{\prime \prime}$ west of the central star.
A30 possesses a large shell ($120 ^{\prime \prime}$) visible  in both 
H$_\alpha$ and OIII, and a hydrogen deficient central region
visible in OIII but not in H$_\alpha$ (Jacoby 1979).
The radio emission emanates from the central region.
The source   extends  to the vicinity of  the main
blobs of the   HST image (Borkowski
{\it et al.} 1995), but it cannot be identified with any of them.
This is compatible with our model where  
the radio emission results from magnetic amplification
in the turbulent region formed by decelerating clumps.

A somewhat parallel situation occurs in  young SNRs. 
In the SNR  Cassiopea A (Cas A), the radio knots as well as the optical 
knots  appear to be connected to the clumpy component of the ejecta,
however, they are not identical. 
Anderson {\it et al.} (1994) analyze the observations of 
optical and radio emission knots in the SNR Cas A.
They claim that while the clumps are almost certainly sites
of particle acceleration, when there is substantial magnetic field
 growth, it will enhance the synchrotron
 emission.
Complex flow patterns bring in particles that were accelerated in
different regions. In young SNRs particles are produced in many acceleration
sites: the blast wave shock, the reverse shock and many internal interactions
 shocks.
Thus the enhanced emission in the radio knots is probably
 a result of field 
amplification. The numerical simulations of Jun \& Jones (1998)
 and  of Jun, Jones \& Norman (1996) 
support this conclusion.

\section {DISCUSSION AND SUMMARY}

 We studied the interaction of a weakly magnetized fast wind, 
of velocity $v_w > 1,000 \km s^{-1}$, blown by the central stars 
of PNs with high density condensations, and  
we propose that it leads to the amplification of the magnetic field.
 When this interaction is with a cluster of clumps,  the magnetic 
field lines are trapped for a long time in the interaction region.
 This process is equivalent to the  numerical simulations of 
Jones {\it et al.} (1996), in which the 2D nature of the 
 flow did not allow the 
magnetic field lines to slide around the condensation.
Jones {\it et al.} (1996) found that the magnetic field can reach 
equipartition with the wind's ram pressure. 
 
 The enhanced magnetic field will have several effects:
(1) It will reduce the mixing (entrainment) of the clump material 
     with the fast wind.
(2) It will reduce the heat conduction between the cool clump and 
      the hot shocked fast wind material.
(3) The magnetic field lines will reconnect among themselves. 
     This process may lead to heating and further X-ray emission.  
(4) It may result in nonthermal radio emission.
The first three processes will affect the X-ray emission from the hot 
material.
 In the current study we did not examine these processes.  
We  explored here, and concluded  that the
expected nonthermal radio emission can be detected in PNs with 
fast winds and  clumps close to the central star. 
 Such a PN is A30 (Dgani {\it et al.} 1998),  claimed also to 
have extended X-ray emission partially overlap with the location of the 
optical condensation (Chu, Chang \& Conway 1997).

 The required initial magnetic field for detectable nonthermal radio 
emission is $\sim 2$ orders of magnitude lower than the field intensity 
required to influence the shape of PNs, according to the model of
CL and Garcia-Segura (1997).
 The stronger magnetic field required by CL and Garcia-Segura (1997)
requires much faster rotation, as shown by equation (2) with 
$\sigma > 10^{-4}$.
This is very problematic  for their model, since
the central stars of PNs are expected to rotate slowly
because of mass loss. This is true  even if their AGB 
progenitors are spun-up by companions on the AGB (Soker 1998). 
 The field in the model of CL and Garcia-Segura (1997)
is globally large. 
 The amplification of the magnetic field through the interaction of the 
magnetized wind with a cluster of condensations, on the other hand,
enhances the field locally, but not globally. 
 Therefore, the detection of strong magnetic fields in specific locations 
in PNs does not necessarily mean that the magnetic field is globally strong.

The process proposed in this study predicts detectable radio 
synchrotron emission from PNs having fast winds and high mass loss rates, 
both of which determine the magnetic field intensity (eq. 7), 
and optical condensations.
 In order to detect this synchrotron emission we need to resolve the 
central region of the nebula from the shell. 
 This requires  high spatial resolution observations.

 To list several PNs, in addition to A30 and
the other hydrogen deficient PNs (see discussion
 in Dgani {\it et al.} 1998) , for which we expect 
nonthermal radio emission we select PNs having fast winds from 
Patriarchi \& Perinotto (1991) and check their optical images for 
condensations in the inner region.  
 We find that from the list of Patriarchi \& Perinotto (1991) 
only NGC 2371 and NGC 7094 show no structure in the very inner regions. 
 Therefore, in general all PNs with fast winds blown by their central 
stars should be searched for nonthermal emission, 
 in particular, those with strong condensations in their central regions:
A78, NGC 5189 and NGC 6210.   
\bigskip

{\bf ACKNOWLEDGMENTS:} 
It is a pleasure to thank Kazik Borkowski, Patrick Harrington, Neal Evans, 
Mordecai Mac-Low, and Stephen White, for useful discussions.
We thank the referee Tom Jones for insightful suggestions.
We thank Sarah Baird,  Neal Evans
and Stephen White for a careful reading of the manuscript and useful comments.
 This research has been supported in part by a grant from the 
Israel Science Foundation.

\end{document}